\documentclass[aps,prl,twocolumn,groupedaddress,showpacs]{revtex4}

\usepackage{graphicx}

\begin{document}

\title{Mapping the magic numbers in binary Lennard-Jones clusters}

\author{Jonathan P.~K.~Doye}
\affiliation{University Chemical Laboratory, Lensfield Road, Cambridge CB2 1EW, United Kingdom} 
\author{Lars Meyer}
\affiliation{University Chemical Laboratory, Lensfield Road, Cambridge CB2 1EW, United Kingdom} 

\date{\today}

\begin{abstract}
Using a global optimization approach that directly searches for the composition
of greatest stability, we have been able to find the particularly stable
structures for binary Lennard-Jones clusters with up to 100 atoms for 
a range of Lennard-Jones parameters. In particular, we have shown
that just having atoms of different size leads to a remarkable
stabilization of polytetrahedral structures, including both polyicosahedral
clusters and at larger sizes structures with disclination lines.
\end{abstract}


\maketitle

The structure of binary clusters has been the subject of much recent interest,
both because of the technological importance of alloy clusters, 
such as in catalysis, and 
the opportunity to tailor the structure through the 
choice of atom types and composition \cite{Sabo04a}, potentially leading 
to novel structural forms, such as the core-shell structures recently found
for 
silver alloy clusters \cite{Rossi04}. Binary clusters
also offer considerable additional challenges to the theoretician, 
compared to the one-component case. Firstly, for a given cluster, there
are many more minima on the potential energy surface, 
because of the presence of ``homotops'' \cite{Jellinek96}, isomers with
the same geometric structure, but which differ in the 
labelling of the atoms.
Secondly, the composition provides an 
additional variable that adds to the complexity of the structural
behaviour.  

For example, 
the task of obtaining 
the lowest-energy structures
for all compositions and all sizes up to 100 atoms would 
require 5050 different global minima to be found.
Our approach here is different, as normally one is not 
interested in all these possible structures, but
only the most stable. Therefore, in our global optimization runs
the composition is allowed to change, and so we attempt to find the
cluster at a given size with the optimal composition directly. 
Thus, the task has been reduced back down to finding one global minimum
for each size,
as for the one-component case. Of course, the search space for each 
optimization is extremely large, and so it is very important to make
extensive use of moves that change the identity of atoms \cite{Calvo04} 
in order to search the space of homotops and different compositions
as efficiently as possible.

The focus of the current work is on
how different structures,
particularly those that are polytetrahedral \cite{NelsonS}, can 
be stabilized just through the two atom types in 
the cluster having different sizes \cite{Rossi04}. 
In polytetrahedral structures 
all the occupied space can be divided up into tetrahedra with 
atoms at their corners. 
However, 
regular tetrahedra
cannot pack all space, and so polytetrahedral packings are
said to be frustrated.
For example, in the 13-atom icosahedron, which can be considered
to be made up of 20 tetrahedra sharing a common vertex, 
the distance between adjacent atoms on the surface is 5.15\% longer
than that between the central atom and a surface atom.
However, the associated strain can be removed by choosing 
the central atoms to be 9.79\% smaller \cite{Cozzini96}. 
Similarly, Frank-Kasper phases,
bulk polytetrahedral crystals, are only found for 
alloys \cite{Shoemaker}.

\begin{figure}
\includegraphics[width=8.4cm]{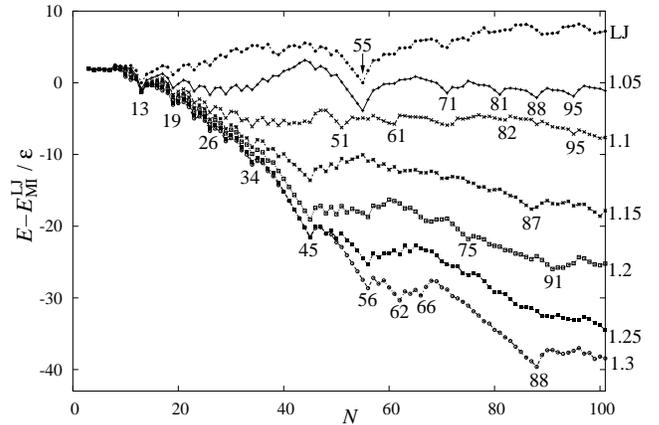}
\caption{\label{fig:eall}The energy of the BLJ global minimum for the
six values of $\sigma_{BB}/\sigma_{AA}$
studied, relative to $E^{\rm LJ}_{\rm MI}$, a fit 
to the energies of the Mackay icosahedra for LJ clusters.
A line corresponding to the LJ global minima is also included
}
\end{figure}

To achieve our aims we use a binary Lennard-Jones (BLJ) potential:
\begin{equation}
E=4\sum_{i<j} \left({\sigma_{\alpha\beta}\over r_{ij}}\right)^{12}-
              \left({\sigma_{\alpha\beta}\over r_{ij}}\right)^6, 
\end{equation}
where $\alpha$ and $\beta$ are the atom types of atoms $i$ and $j$,
respectively.
To study the effects of different 
atom sizes, independent of energetic effects, 
we choose 
$\epsilon_{AA}=\epsilon_{AB}=\epsilon_{BB}=\epsilon$
and define $\sigma_{AB}$ using the Lorentz rule:
$\sigma_{AB}=(\sigma_{AA}+\sigma_{BB})/2$. 
The one parameter
in the potential is then $\sigma_{BB}/\sigma_{AA}$.
For this choice of parameters a tendency to form core-shell
clusters has been observed \cite{Garzon89,Clarke94}, 
but no systematic structural survey has been made.
Our aim here is to find how the stable structures of the BLJ
clusters change, as $\sigma_{BB}/\sigma_{AA}$ varies in the 
range 1.0 to 1.3 for all clusters with up to 100 atoms.

\begin{figure}
\includegraphics[width=8.4cm]{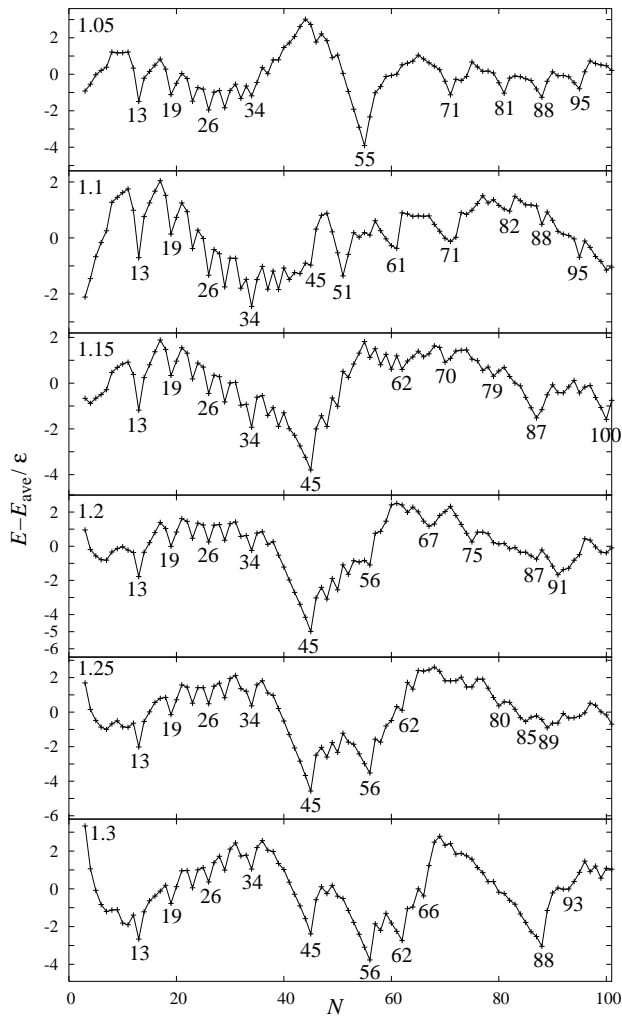}
\caption{\label{fig:enorm}The six panels correspond to the
energy of the global minimum for the
six values of $\sigma_{BB}/\sigma_{AA}$
studied, relative to $E_{\rm ave}$, a fit to the 
energies of the global minima at that size ratio.}
\end{figure}

\begin{figure}
\includegraphics[width=8.4cm]{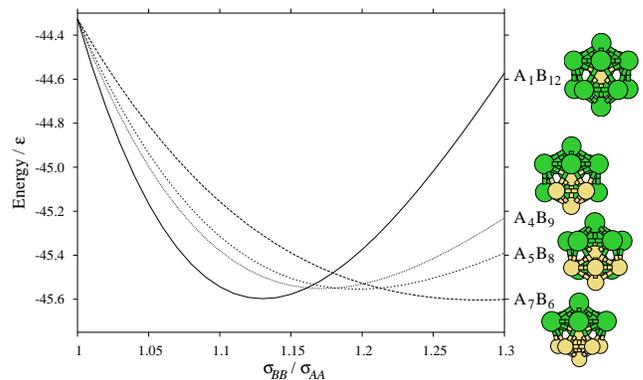}
\caption{\label{fig:BLJ13} (Colour online)
The dependence of the energies of the four BLJ$_{13}$ global minima on
$\sigma_{BB}/\sigma_{AA}$.
}
\end{figure}

\begin{figure*}
\includegraphics[width=18cm]{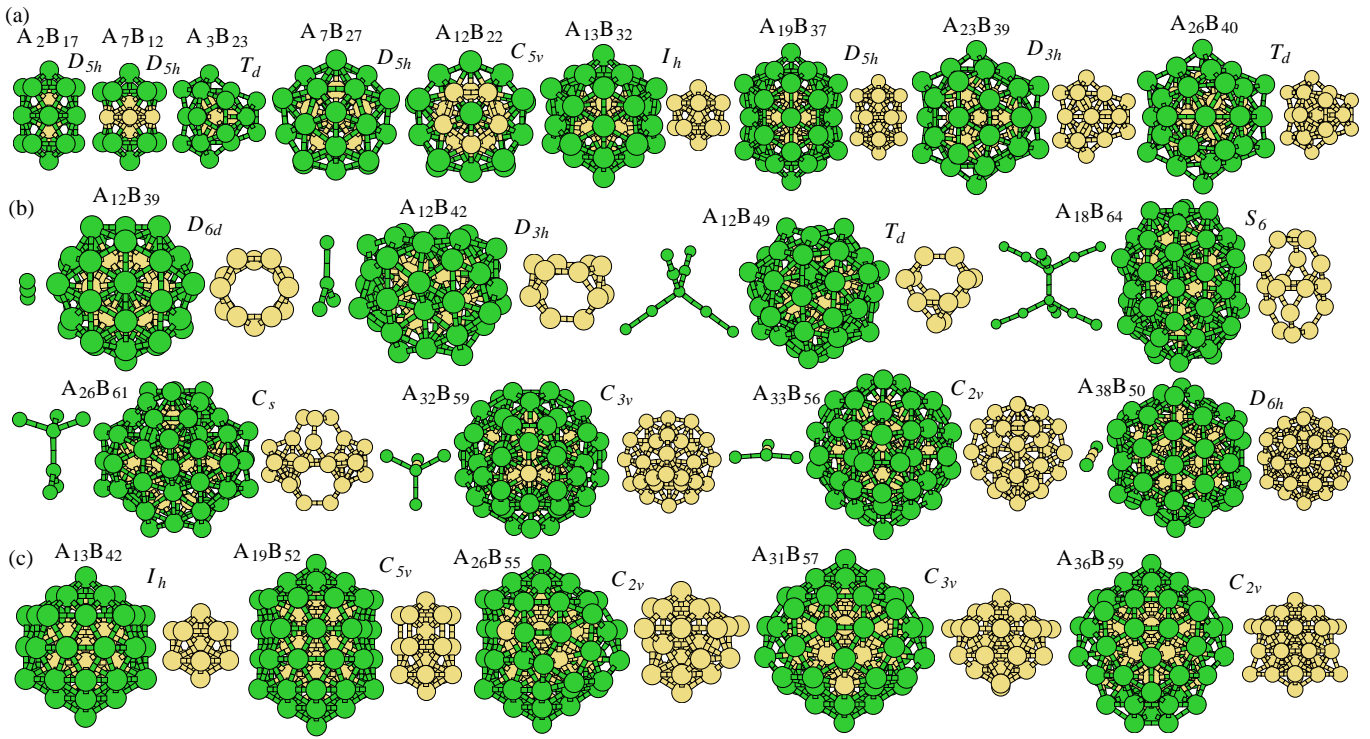}
\caption{\label{fig:structures} (Colour online)
A selection of the particularly stable BLJ global minima grouped
according to their structural type: (a) polyicosahedral, (b) polytetrahedral
with disclinations and (c) the 55-atom Mackay icosahedron with an anti-Mackay
overlayer.
For the larger clusters, to the right of the cluster, the A-atom core is also
depicted, and in (b) to the left is the disclination network.
}
\end{figure*}

\begin{figure}
\includegraphics[width=8.4cm]{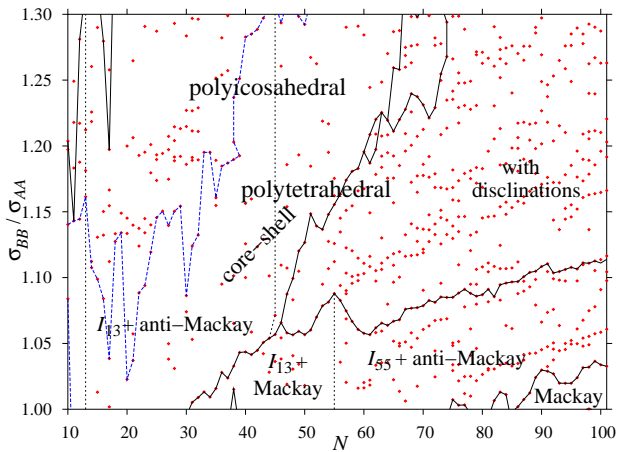}
\caption{\label{fig:phased} (Colour online)
Structural phase diagram showing how the structure
of the global minimum depends on $N$ and 
$\sigma_{BB}/\sigma_{AA}$. Each point corresponds to the
value of $\sigma_{BB}/\sigma_{AA}$ at which the global minimum
for a given $N$ changes. The lines divide the diagram into 
regions where the global minima have the same structural type. 
The labels $I_{13}$ and $I_{55}$ stand for the 13- and 55-atom 
Mackay icosahedra.}
\end{figure}

The energies of the putative global minima that we have found
are depicted in Figures \ref{fig:eall} and \ref{fig:enorm}, 
where Fig.\ \ref{fig:eall} compares the energies at different values
of $\sigma_{BB}/\sigma_{AA}$ and Fig.\ \ref{fig:enorm} allows 
the magic numbers at each size ratio to be identified more easily.
Figure \ref{fig:BLJ13} provides a more detailed analysis of the 
behaviour of BLJ$_{13}$,
Figure \ref{fig:structures} shows a selection of particularly 
stable structures and 
Figure \ref{fig:phased} how the structural form of the 
global minima depends on $N$ and $\sigma_{BB}/\sigma_{AA}$. 
The energies and points files for all
the global minima are available online \cite{CCDshort}.

The reference system to which our results are compared is the 
one-component Lennard-Jones (LJ) clusters, for which the structural
behaviour is well-understood. 
In the present size range, the LJ global minima
are dominated by structures based upon the Mackay icosahedra \cite{Northby87}. 
These Mackay icosahedra are made up of twenty face-centred-cubic (fcc) 
tetrahedra sharing a common vertex, and except for the smallest icosahedron at 
$N=13$ are not polytetrahedral. 

Growth upon the Mackay icosahedra can occur in two ways. The first, the 
anti-Mackay overlayer, consists in adding atoms in sites that are hexagonal
close-packed with respect to the underlying fcc tetrahedra and above the
twelve vertices. 
For the 13-atom icosahedron, this overlayer maintains the
polytetrahedral character of the clusters. 
The second, the Mackay overlayer continues the fcc packing 
of the underlying tetrahedra and leads to the next Mackay icosahedron.
Growth initially occurs in the anti-Mackay overlayer
because of a greater number of nearest-neighbour contacts,
but before this overlayer is complete,
the LJ global minimum changes to Mackay character \cite{Northby87}, 
because of the greater strain energy associated with the anti-Mackay overlayer.

It is immediately clear from Fig.\ \ref{fig:eall} that
allowing atoms of different sizes leads to a dramatic stabilization of 
the clusters. 
For example, the BLJ$_{45}$ global minimum at $\sigma_{BB}/\sigma_{AA}$=1.3
is $26.9\,\epsilon$ or 12.6\% lower in energy than that for LJ$_{45}$.
The origins of this stabilization are quickly apparent 
from an analysis of the structural behaviour of the BLJ clusters. 
The structural phase diagram in Fig.\ \ref{fig:phased} shows
that for the majority of the parameter space, the global minima
are polytetrahedral. Only in the bottom right-hand corner (large $N$ and
low size ratio) are non-polytetrahedral structures most stable.
For LJ clusters, such polytetrahedral structures are disfavoured beyond 30 atoms
because of their greater strain energy, however the presence of
different-sized atoms relieves this strain. 
As the polytetrahedral structures generally 
have a greater number of nearest neighbours, they therefore
become lowest in energy.

Perhaps surprisingly, there is no optimal size ratio, but instead
the energy virtually monotonically decreases 
with increasing $\sigma_{BB}/\sigma_{AA}$ (Fig.\ \ref{fig:eall}).
For example, based on the analysis of the 
geometry of the icosahedron mentioned earlier, one might 
expect a size difference near 10\% to be optimal for BLJ$_{13}$. 
Indeed, the energy of A$_1$B$_{12}$ 
shows a pronounced minimum near this value (Fig.\ \ref{fig:BLJ13}), 
however just beyond this minimum the optimal structure changes.
The low energy is maintained by adding an increasing number
of the smaller A atoms into the surface of the icosahedron, as this 
keeps the nearest-neighbour distances near to their optimal values.
A similar story holds for larger clusters, except that 
the optimal values of $\sigma_{BB}/\sigma_{AA}$ for the core-shell geometry
are larger (Fig.\ \ref{fig:phased})
because
greater size ratios are needed to fully relieve the strain.

This preference for polytetrahedral structures is evident in the magic 
numbers  (Fig.\ \ref{fig:enorm}).  Only for $\sigma_{BB}/\sigma_{AA}$=1.05 
is the 55-atom Mackay icosahedron still a magic number.
Instead the magic numbers 
at $N$=19, 23, 26, 29, 34, and 45 
associated with 
the covering of the 13-atom icosahedron by the anti-Mackay overlayer 
become increasingly prominent.
These structures are polyicosahedral---each atom in the interior of the 
cluster has a local icosahedral coordination shell---and 
are made up of 2, 3, 4, 5, 7 and 13 interpenetrating icosahedra, respectively
(Fig.\ \ref{fig:structures}).
Recently, similar polyicosahedral core-shell structures have been
found for alloy clusters of silver \cite{Rossi04}.
Although a significant proportion of the polyicosahedral region
of the structural phase diagram corresponds to core-shell clusters
(Fig.\ \ref{fig:phased}),
the stability of these structures is not dependent on
such an arrangement, and as for BLJ$_{13}$, A atoms are incorporated into
the surface at larger $\sigma_{BB}/\sigma_{AA}$,
as illustrated by A$_7$B$_{12}$ and A$_{12}$B$_{22}$ in 
Fig.\ \ref{fig:structures}.

The complete anti-Mackay covering of the 13-atom icosahedron occurs at
$N$=45, however at larger size ratios the polyicosahedral growth
continues beyonds this size. The magic numbers at $N$=56, 62 and 66 
correspond to core-shell polyicosahedral structures with the
double, triple and quadruple icosahedra mentioned above as their core 
(Fig.\ \ref{fig:structures}).

As $N$ increases the polyicosahedral structures have increasingly large 
tensile strains in the surface,
hence the need for increasingly large size ratios to 
counteract this.
Furthermore, polyicosahedral structures are impossible
for bulk. Instead, the Frank-Kasper phases also involve coordination 
numbers greater than 12. Such structures can be described using 
disclinations, where a disclination runs along those edges in the 
structure that have six tetrahedra surrounding them, rather than the
usual five for icosahedral coordination \cite{NelsonS}.
Polytetrahedral clusters involving disclinations introduce 
both tensions and compressions into the structure, and represent
a better compromise at larger $N$.
Indeed, such structures cover a significant proportion of the 
phase diagram at larger $N$ (Fig.\ \ref{fig:phased}),
and some examples are illustrated in 
Fig.\ \ref{fig:structures}. 

Unlike the polyicosahedral structures,
the core of the cluster is not necessarily made up completely 
of A atoms, but instead the atoms with coordination number ($Z$) greater
than 12 usually correspond to the larger B atoms.
Atoms with $Z$=14 have a single disclination running
through them, whilst atoms with $Z$=15 and 16 are nodes
for three and four disclinations, respectively.

Some of the polytetrahedral magic numbers at $\sigma_{BB}/\sigma_{AA}=1.1$
have been previously seen for one-component clusters interacting with
long-ranged Morse \cite{Doye97d} and 
modified Dzugutov \cite{Doye01d} potentials.
The structures at $N$=51, 54 and 61 consist of 
the $Z$=14, 15 and 16 coordination polyhedron, respectively, 
covered by a (near-)complete anti-Mackay-like overlayer.
As such, they are the high coordination number analogues
of A$_{13}$B$_{32}$. Similarly, A$_{18}$B$_{64}$ is the analogue of 
A$_{19}$B$_{37}$, but with two interpenetrating Z=16 coordination polyhedra
at the centre.

As $\sigma_{BB}/\sigma_{AA}$ increases, the fraction of the atoms
that lie on disclination lines decreases,
reaching zero at the polyicosahedral boundary in the structural phase
diagram. This trend is illustrated by the four structures in the second
line of Fig.\ \ref{fig:structures}(b), which correspond to magic numbers
for $\sigma_{BB}/\sigma_{AA}$=1.15, 1.2, 1.25 and 1.3, respectively.
For example, the hexagonal disk structure that occurs for 
$\sigma_{BB}/\sigma_{AA}$=1.3 at $N$=88, has a single disclination running
along the six-fold symmetry axis, and its A-atom core is a 38-atom 
structure previously seen for Dzugutov clusters \cite{Doye01a,Doye01d}.

The major non-polytetrahedral portion of the structural phase diagram
corresponds to structures based on the 55-atom Mackay icosahedron.
As for growth on the 13-atom icosahedron, the size at which the
transition from an anti-Mackay to a Mackay overlayer occurs increases with
increasing $\sigma_{BB}/\sigma_{AA}$. At $\sigma_{BB}/\sigma_{AA}$=1.05, this
transition does not occur in the present size range, even though it 
begins at $N$=82 for LJ clusters. Fig.\ \ref{fig:structures}(c) illustrates
some of the magic numbers with an anti-Mackay overlayer that occur for 
$\sigma_{BB}/\sigma_{AA}$=1.05 and 1.1, all of which have a core-shell 
geometry. 

In summary, we have developed a global optimization approach for binary
clusters that is able to locate magic number clusters up to 
unprecedented sizes, and which should also prove particularly useful for 
analysing bimetallic clusters. We have applied this approach
to binary Lennard-Jones clusters, our hope being that, in the same way as 
for Lennard-Jones clusters in the one-component case, this system will
become a simple archetypal system both to provide candidate structures
for a wide variety of binary clusters and to rationalize their structures.
Here, we focussed on the case where only the sizes of the two atom types are 
different. This leads to a remarkable stabilization of polytetrahedral 
clusters, and a zoo of interesting structures.
Because clusters have been found to provide a good indicator
of the preferred local structure within supercooled liquids 
\cite{Frank52,Doye01a},
these results can also provide an interesting perspective on the role of 
size mismatch on glass formation in binary systems. Not only does size mismatch
enhance glass formation due to the destabilization of a crystalline solid
solution \cite{Egami84}, but we see here that it also encourages local 
icosahedral coordination \cite{Lee03b}, hence frustrating crystallization 
further.

\begin{acknowledgments}
J.P.K.D is grateful to the Royal Society for financial support.
\end{acknowledgments}

\end{document}